\documentclass[11pt,reqno]{amsart}

\usepackage{amsmath,amsfonts,amsthm,amscd,amssymb,graphicx,mathrsfs}
\usepackage[utf8]{inputenc}
\usepackage{amsbsy}
\usepackage{graphicx}
\usepackage{subcaption}
\usepackage{hyperref}
\usepackage{bm}
\usepackage{tikz}
\usetikzlibrary{arrows}
\usepackage[text={6.5in,9in},centering,includefoot,foot=0.6in]{geometry}

\usepackage[all]{xy}

\usepackage{amsmath,amsthm,amsfonts,amssymb,amscd}
\usepackage[all]{xy}
\usepackage{sseq}

\usepackage{todonotes}

\definecolor{skyblue}{rgb}{0.85,0.85,1}

\newtheorem{theorem}{Theorem}[section]

\newtheorem{rem}[theorem]{Remark}
\newtheorem{define}[theorem]{Definition}

\newcommand{\p}{\partial}				

\DeclareMathOperator{\dv}{div}

\DeclareMathOperator{\tr}{tr}

\newcommand{\cM}{{\mathcal{M}}}
\newcommand{\cS}{{\mathcal{S}}}
\newcommand{\cT}{{\mathcal{T}}}

\newcommand{\R}{\mathbb{R}}

\newcommand{\overbar}[1]{\mkern 1.5mu\overline{\mkern-2.5mu#1\mkern-1.5mu}\mkern 1.5mu}

\begin{document}

\title{Eigenvalues of the MOTS stability operator for slowly rotating Kerr black holes}
\author[L. Bussey]{Liam Bussey}\email{liamb@mun.ca}
\author[G. Cox]{Graham Cox}\email{gcox@mun.ca}
\author[H. Kunduri]{Hari Kunduri}\email{hkkunduri@mun.ca}\address{Department of Mathematics and Statistics, Memorial University of Newfoundland, St. John's, NL, Canada}

\maketitle
\begin{abstract}
We study the eigenvalues of the MOTS stability operator for the Kerr black hole with angular momentum per unit mass $|a| \ll M$. We prove that each eigenvalue depends analytically on $a$ (in a neighbourhood of $a=0$), and compute its first nonvanishing derivative. Recalling that $a=0$ corresponds to the Schwarzschild solution, where each eigenvalue has multiplicity $2\ell+1$, we find that this degeneracy is completely broken for nonzero $a$. In particular, for $0 < |a| \ll M$ we obtain a cluster consisting of $\ell$ distinct complex conjugate pairs and one real eigenvalue. As a special case of our results, we get a simple formula for the variation of the principal eigenvalue. For perturbations that preserve the total area or mass of the black hole, we find that the principal eigenvalue has a local maximum at $a=0$. However, there are other perturbations for which the principal eigenvalue has a local minimum at $a=0$.
\end{abstract}

\section{Introduction}
A classic result of general relativity states that the two-parameter family of Kerr metrics  exhausts the set of all asymptotically flat, stationary and axisymmetric black hole solutions of the vacuum Einstein equations (see \cite[Theorem 3.2]{Chrusciel:2012jk} for precise hypotheses).  This remarkable result asserts that the spacetime outside an isolated rotating astrophysical black hole at equilibrium should be describable to sufficient accuracy by the Kerr solution.  On the other hand,  most physical processes of interest involving black holes are dynamical. Asserting the existence of an event horizon requires full knowledge of the global time evolution of the spacetime, which is a difficult problem even in highly symmetric settings.  Accordingly,  a quasi-local characterization of black holes, which can be stated within the initial value formulation of the Einstein equations,  is required to model such phenomena as the merger of two black holes. 

Such a characterization is provided by the notion of a marginally outer trapped surface (MOTS) \cite{Andersson:2005gq, Andersson:2007fh}. Given a codimension two closed spacelike surface $\cS$, one considers the null expansion $\theta_+$ of future pointing outgoing geodesics emanating from $\cS$.  Such a surface is outer trapped if $\theta_+ <  0$. A MOTS is characterized by $\theta_+ =0$. Importantly, this definition is consistent with the familiar stationary case: a spatial cross section of the event horizon of a stationary black hole is indeed a MOTS.  

Given a foliation of spacetime into spacelike hypersurfaces $\Sigma_t$, suppose that initially at $t=0$, $\Sigma_0$ contains a MOTS $\cS_0$. The authors of \cite{Andersson:2008up} proved that if $\cS_0$ satisfies a certain stability condition, then the MOTS will propagate in time into a marginally outer trapped tube, whose marginally outer trapped leaves $\cS_t$ lie  in $\Sigma_t$. This stability condition can be intuitively thought of as a Lorentzian analogue of the notion of stability of a minimal surface.   Associated to a given MOTS $\cS$ lying in a hypersurface $\Sigma$, one may consider the linearization  $L$ of the expansion $\theta_+$ under smooth variations of $\cS$ generated by normal deformations lying within $\Sigma$ ($L$ is defined in \eqref{L} below).  $L$ is a linear, second-order, elliptic differential operator on $\cS$. In general it is not self-adjoint, and hence may have complex eigenvalues. However, its principal eigenvalue $\lambda_{\rm p}$ (defined as the eigenvalue with smallest real part) is known to be real. The MOTS $\cS$ is then said to be stable if $\lambda_{\rm p} \geq0$ and strictly stable if $\lambda_{\rm p} > 0$.

The stated result of \cite{Andersson:2008up} is that $\cS$ can be smoothly propagated in time if it is strictly stable, but the same conclusion holds under the significantly weaker assumption that $0$ is not an eigenvalue of $L$. (Strict stability implies \emph{all} eigenvalues of $L$ have positive real part, and hence are nonzero.) The importance of this generalization is made clear by recent numerical simulations of black hole mergers \cite{Pook-Kolb:2020zhm,PBJKS20,PBKS19};  these contain MOTS for which the principal eigenvalue is negative, hence one needs to check whether or not any of the higher eigenvalues vanish. It is also possible to give analytic examples where the principal eigenvalue is negative (such as the Reissner--Nordstr\"om spacetime for certain parameter values, see \cite{BKO17}), so the time propagation of the MOTS again may depend on the higher eigenvalues of $L$.
Furthermore, it has been suggested, via an analogy with fluid dynamics, that higher eigenvalues are related to dynamical stability~\cite{Jaramillo:2013rda,J15}. Further discussion on the evolution of horizons can be found in \cite{Booth:2006bn}.

In general, eigenvalues of such differential operators cannot be computed explicitly, except in highly symmetric situations, such as \cite{J15,J15b,K19}. In particular, the lack of self-adjointness means that many basic tools of spectral theory, such as the Courant min-max principle, are not available. We mention in passing the intriguing ``analyticity conjecture" of  \cite{J15b}, which proposes that the eigenvalues of $L$ can be obtained by analytic continuation from an associated self-adjoint operator.

In this paper we study the spectrum of the Kerr MOTS. The Kerr solution is characterized by its ADM mass $M$ and the angular momentum per unit mass $a = J / M \in [-M,M]$, where $J$ is the ADM angular momentum. We are interested in the eigenvalues of the stability operator, $L(a)$, for small values of $a$. We compute these perturbatively, recalling that when $a=0$, the Kerr solution reduces to the spherically symmetric Schwarzschild solution. In this case the stability operator simplifies to
\begin{equation}\label{LSchw}
	L(0) = - \frac{1}{4M^2} \Delta_{S^2} + \frac{1}{4M^2},
\end{equation}
where $\Delta_{S^2}$ is the Laplace--Beltrami operator on the two-sphere, and so the eigenvalues are given by
\begin{equation}
\label{spec:unperturbed}
	\lambda_\ell = \frac{\ell(\ell + 1) + 1}{4M^2}
\end{equation}
for $\ell = 0, 1, 2, \ldots$. Each eigenvalue $\lambda_\ell$ has multiplicity $2\ell + 1$, with an eigenbasis given by the spherical harmonics $Y_\ell^m(\theta,\phi)$ for $-\ell \leq m \leq \ell$.

While it is natural to fix $M$ and vary $a$, it is also interesting to consider variations that preserve the area of the horizon. To do this we must allow the mass to depend on $a$. 
The black hole occurs at radius $r_+ = M + \sqrt{M^2 - a^2}$ and has area $A = 4\pi(r_+^2 + a^2)$. We will see below that the stability operator can be written in terms of $a$ and $r_+(a)$, with $M$ not appearing explicitly, so the variation of mass is encoded in the function $r_+(a)$. While our approach works for any variation $r_+(a)$, we emphasize the following examples:
\begin{align*}
	\text{fixed mass} & \ \Longleftrightarrow \ r_+(a) = \overbar M + \sqrt{\overbar M ^2 - a^2} \\
	\text{fixed area} & \ \Longleftrightarrow \ r_+(a) = \sqrt{4 \overbar M^2 - a^2} 
\end{align*}
These both have $r_+(0) = 2 \overbar M$ and $r_+'(0) = 0$, where $\overbar M$ denotes the unperturbed mass.

\begin{theorem}\label{thm:HF}
Suppose $r_+(a)$ is analytic in a neighbourhood of $a=0$, with $r_+(0) = 2 \overbar M > 0$ and $r_+'(0) = 0$.
Fix $\ell  \geq 0$. In a neighbourhood of $a=0$ there exist analytic curves of eigenvalues $\lambda^m_\ell(a)$, $-\ell \leq m \leq \ell$, with corresponding eigenfunctions  $\psi^m_\ell(a)$, such that $\lambda_\ell^m(0) = \lambda_\ell$, $\psi_\ell^m(0) = Y_\ell^m$, and
\begin{equation}\label{eq:lambdalm}
	(\lambda_\ell^m)'(0) = \frac{3m}{8 \overbar M^3} i.
\end{equation}
\end{theorem}

Therefore, $\lambda_\ell$ splits completely into $2\ell+1$ distinct eigenvalues, consisting of one real eigenvalue and $\ell$ complex conjugate pairs. The rate at which each conjugate pair moves away from the real axis is proportional to the magnetic quantum number $m$, but does not depend on $\ell$. In particular, the $m=0$ eigenvalue branch is constant to first order in $a$. The second derivative, however, does not vanish in general.

\begin{theorem}\label{thm:2nd}
The eigenvalue curve $\lambda_\ell^0(a)$ is real in a neighbourhood of $a=0$, with
\begin{align}\label{eq:lambdal}
	(\lambda_\ell^0)''(0) = - \frac{4 \overbar M r_+''(0) + 5}{16 \overbar M^4} + \frac{\ell(\ell+1)}{4 \overbar M^4}
	\left( \frac{\ell(\ell+1) - \frac54}{(2\ell+3)(2\ell-1)} - \frac{2 \overbar M r_+''(0) + 1}{2} \right).
\end{align}

\end{theorem}

The principal eigenvalue of $L(0)$ is $\lambda_{\rm p} = \lambda_0 = 1/(4 \overbar M^2)$. Since the eigenvalues $\lambda_\ell^m(a)$ depend continuously on $a$, it is guaranteed that $\lambda_0^0(a)$ will be the principal eigenvalue of $L(a)$ for sufficiently small $a$.
We therefore obtain
\begin{equation}
\label{prin:var}
	\lambda_{\rm p}'(0) = 0, \qquad \lambda_{\rm p}''(0) = - \frac{4 \overbar M r_+''(0) + 5}{16 \overbar M^4},
\end{equation}
as a special case of Theorems \ref{thm:HF} and \ref{thm:2nd}. For the different choices of $r_+(a)$ discussed above, we obtain
\begin{equation}
	\lambda_{\rm p}''(0) = \begin{cases} - \frac{1}{16 \overbar M^4}, & \text{fixed mass} \\
	- \frac{3}{16 \overbar M^4}, & \text{fixed area} 
	\end{cases}
\end{equation}
For these variations we conclude that the principal eigenvalue is a concave function for small values of $a$, and in particular has a local maximum at $a=0$. On the other hand, for any variation with $r_+''(0) < -\frac{5}{4 \overbar M}$, the principal eigenvalue will have a local \emph{minimum} at $a=0$.

The variation of the principal eigenvalue with fixed mass was obtained in the first author's B.Sc. Honours thesis \cite{B20}. The variation with $r_+(a) = 2 \overbar M$ fixed was computed in \cite{thesis2} and found to be $- \frac{5}{16 \overbar M^4}$, which agrees with \eqref{prin:var} when $r_+''(0) = 0$. The first variation was computed formally in \cite{thesis1}; that is, it was shown that \emph{if} the eigenvalues are differentiable, then their derivatives must be given by~\eqref{eq:lambdalm}.

Before proving the results, we make a few remarks, hinting at the proofs and some of the difficulties encountered therein.

\begin{rem}
The analyticity of the eigenvalues and eigenfunctions in Theorem \ref{thm:HF} is nontrivial, since the unperturbed eigenvalue $\lambda_\ell$ is degenerate and $L(a)$ is not self-adjoint for $a \neq 0$. The key to the proof is the fact that the degeneracy is completely broken at first order in $a$, i.e. the $2\ell+1$ numbers appearing in \eqref{eq:lambdalm} are distinct.
\end{rem}

\begin{rem}
The first derivative of $L(a)$ at $a=0$ only contains a single term, proportional to $\p /\p\phi$. This greatly simplifies the computation of \eqref{eq:lambdalm}, and explains why $\lambda_\ell^0(a)$ is constant to first order in $a$, as the corresponding eigenfunction $\psi_\ell^0(a) = Y_\ell^0 + \mathcal{O}(a)$ does not depend on $\phi$ at zeroth order in $a$.
\end{rem}

\begin{rem}
To prove Theorem \ref{thm:2nd} we will show that $\lambda_\ell^0(a)$ is the $\ell$th eigenvalue of a singular Sturm--Liouville problem, on a space of functions depending only on $\theta$. For this reduced problem $\lambda_\ell^0(a)$ is a simple eigenvalue. Moreover, the reduced stability operator depends only on $a^2$, so we only need to differentiate once in $a^2$ to obtain \eqref{eq:lambdal}.
\end{rem}

\section*{Acknowledgments}
The authors would like to thank Jos\'e Luis Jaramillo for helpful comments and discussions on this problem. G.C. and H.K.  acknowledge the support of NSERC grants RGPIN-2017-04259 and RGPIN-2018-04887, respectively. 

\section{The stability operator on cross sections of the Kerr event horizon}

In this preliminary section we define MOTS and the corresponding stability operator, then recall the Kerr black hole solution and its MOTS, culminating in the explicit formula \eqref{Kerr stab} for the stability operator $L(a)$ on a particular cross section of the horizon.

While the stability operator depends explicitly on the cross section, and a choice of spacelike hypersurface containing it, its spectrum is independent of the choice of hypersurface, by \cite[Lemma~1]{J15}. More precisely, for hypersurfaces $\Sigma$ and $\widehat\Sigma$ both containing the surface $\cS$, the resulting stability operators will be related by $\widehat L \psi = f L(f^{-1}\psi)$ for some smooth positive function $f$, and hence have the same spectrum. In general the spectrum \emph{will} depend on the cross section, but in the present situation, \cite[Proposition 4]{Mars:2012sb} implies that the spectrum is actually independent of the choice of cross section, and hence can be viewed as a property of the horizon itself.

\subsection{The stability operator}
	Consider a four-dimensional spacetime $(\cM,g)$ and a spacelike hypersurface $\Sigma$ lying in $\cM$ with unit timelike normal field $n$, induced metric $\gamma$ and second fundamental form $K$. The triple $(\Sigma, \gamma, K)$ constitute an initial data set.  Let $\cS$ be a closed 2-surface embedded in $\Sigma$ with unit spacelike normal $s$ (note $s$ is tangent to $\Sigma$). In the case of interest, $\Sigma$ will have an  asymptotically flat end and $\cS$ divides $\Sigma$ into an `outside' and `inside' with respect to this designated end. Using these two normals, we construct unit null outward and inward pointing vector fields $l_\pm=n\pm s$.
	\begin{define}
		The \textbf{null expansion scalars} of $\cS$, denoted $\theta_\pm$, are the divergence of outgoing and ingoing light rays emerging orthogonally from $\cS$, and hence take the form $\theta_\pm=\dv_{\cS} l_\pm$.
	\end{define}
	Using our initial data set $(\Sigma,\gamma,K)$, we can express the null expansions in the form (see, e.g. the exposition \cite{Galloway:2011np})
	\begin{equation}
		\theta_\pm = \tr_qK\pm H
	\end{equation}
	where $\tr_qK$ is the trace of the extrinsic curvature tensor with respect to $q$, the induced metric on $\cS$, and $H$ is the mean curvature of $\cS$ in $\Sigma$. Note that the null expansions of $\cS$ can be expressed in terms of initial data alone.  The surface $\cS$ is {\it{trapped}} if both $\theta_+ <0$ and $\theta_- < 0$, {\it outer trapped} if $\theta_+ < 0$, and {\it marginally outer trapped} (MOTS) if $\theta_+ = 0$.  All quantities are understood to be evaluated on $\cS$. 

Let $\cS_t$ denote the one-parameter family of surfaces created by deforming a MOTS $\cS$ an amount $t\psi$ in the outward normal direction, where $\psi:\cS\to\R$ is a smooth function. That is,
\begin{equation}
	\cS_t = \big\{ \exp_x\big(t \psi(x) s_x \big) : x \in \cS \big\}
\end{equation}
where $\exp$ is the exponential map for $\Sigma$ and $s_x \in T_x \Sigma$ is the outward unit normal to $\cS$ at $x$. The stability of $\cS$ is studied by investigating the change of $\theta_+$ along these deformations of $\cS$. A computation shows that
	\begin{equation}
		\frac{\p\theta_+}{\p t}\bigg|_{t=0} = L\psi,
	\end{equation}
	where $L$ is the differential operator \cite{Andersson:2007fh}
	\begin{equation} \label{L}
		L\psi = -\Delta_q\psi + 2q(X,\p\psi) + \left(\frac{R_q}{2}-\cT(n,l_+)-\frac{1}{2}|\chi_+|^2+\dv X-q(X,X)\right)\psi.
	\end{equation}
	In the above, $\Delta_q$ and $R_q$ are respectively the Laplacian  and scalar curvature of $(\cS, q)$, $\cT$ is the spacetime stress-energy tensor, $X$ is the vector field on $\cS$ obtained by raising the index on the one-form $X_i=K_{ij}s^j\big|_{\cS}$, and $\chi_+$ is the null second fundamental form tensor with components tangential to $\cS$, obtained from decomposing the second fundamental form of $\Sigma$ and given by
	\begin{equation}
		\chi^\pm_{ab} = \nabla_al_b^\pm.
	\end{equation}
For a stationary spacetime satisfying the null energy condition, such as Kerr, it follows from the Raychaudhuri equation that $\chi^+ = 0$.

\subsection{The Kerr black hole} As is well known, the exterior Kerr spacetime $(\mathbb{R}_{>0} \times \mathbb{R} \times S^2,g)$ is a vacuum solution of the Einstein equations.  We are mainly concerned with the domain of outer communications (the $\mathbb{R}_{>0}$  factor corresponds to a `radial' direction). In Boyer--Lindquist coordinates $(t,r,\theta, \phi)$ the metric is given by
	\begin{equation}
		\begin{aligned}
			g  =& -\frac{(\Delta - a^2 \sin^2\theta)}{\rho^2} dt^2 - 2 a \sin^2\theta \frac{(r^2 + a^2 - \Delta)}{\rho^2} dt d\phi \\
			& + \left(\frac{(r^2 + a^2)^2 - \Delta a^2 \sin^2\theta}{\rho^2} \right) \sin^2\theta d\phi^2 + \frac{\rho^2}{\Delta} dr^2 + \rho^2 d\theta^2,
		\end{aligned}
	\end{equation}
	where we defined the functions
	\begin{equation}
		\rho^2 = r^2+a^2\cos^2\theta, \qquad \Delta = r^2-2Mr+a^2,
	\end{equation}
	with $M\ge0$ and $|a| \le M$. 
	The exterior region is covered by the coordinate ranges $t \in \mathbb{R}, r > r_+$ and $(\theta, \phi)$ are standard coordinates on $S^2$. The spacetime contains a smooth event horizon at  $r=r_+$, where $r_+ = M + \sqrt{M^2 - a^2}$ is the largest positive root of $\Delta$. 

This event horizon is a Killing horizon with associated null generator \begin{equation} \xi = \partial_t + \Omega_H \partial_\phi, \qquad \Omega_H = \frac{a}{r_+^2 + a^2}. \end{equation} 
To obtain an explicit expression for the stability operator, it is useful to introduce a canonical coordinate system adapted to the initial data. We can decompose the spacetime metric $g$ using the $3+1$ formulation, separating the temporal and spatial components as follows
	\begin{equation}
		ds^2 = -\alpha^2dt^2+\gamma_{ij}(dx^i+N^idt)(dx^j+N^jdt),
	\end{equation}
	where $\alpha$, $N^i$, and $\gamma_{ij}$ represent the lapse function, shift vector, and induced metric on $\Sigma$, respectively.  In particular, $\Sigma$ corresponds to a surface $t =0$ in this system.  One easily reads off
	\begin{equation}
\alpha = \left(\frac{\Delta \rho^2}{(r^2+a^2)^2 - \Delta a^2 \sin^2\theta} \right)^{1/2}, \qquad 	N  = -\frac{2 M a r}{(r^2+a^2)^2 - \Delta a^2 \sin^2\theta}  \partial_\phi.
	\end{equation}
	 The extrinsic curvature tensor of $\Sigma$ is related to the time rate of change of the metric using the following expression:
	\begin{equation}
		K_{ij} = \frac{1}{2\alpha}(D_iN_j+D_jN_i-\p_t\gamma_{ij}),
	\end{equation}
	where $D$ denotes the covariant derivative with respect to the metric $\gamma_{ij}$.  The above chart is degenerate at $r =r_+$, as $g_{rr}$ diverges.  By defining a new radial coordinate $\hat{r}$ implicitly by
	\begin{equation}
	r = G(\hat{r}) = \hat{r} + M + \frac{M^2 - a^2}{4\hat{r}}
	\end{equation} we find the induced metric on spatial hypersurfaces is given by
	\begin{equation}
	\gamma = \frac{\rho^2}{\hat{r}^2} (d \hat{r}^2 + \hat{r}^2 d\theta^2) + \left(\frac{(r^2+a^2)^2 - \Delta a^2 \sin^2\theta}{\rho^2}\right) \sin^2\theta d\phi^2.
	\end{equation} The above form of the metric is manifestly regular at the event horizon, which corresponds to $\hat{r} = \hat{r}_+ = \sqrt{M^2 - a^2}/2$.  The original `exterior' region is described by $\hat{r} > \hat{r}_+$ and the metric can be extended to  $\hat{r} < \hat{r}_+$ to reveal a new asymptotically flat region as $\hat{r} \to 0^+$.  Hence $(\Sigma, \gamma, K)$ constitutes a complete vacuum initial data set on $\mathbb{R} \times S^2$ with two asymptotically flat ends.  The notion of  `inward' and 'outward' expansions are understood to be with respect to the original asymptotically flat end  corresponding to $\hat{r} \to \infty$.

It is easily checked, using appropriately normalized normals proportional to $dt$ and $dr$, that the 
two-sphere $\cS$ defined by $\hat r= \hat r_+$ is a MOTS, that is $\theta_+ =0$.  The induced metric on $\cS$ is given by
\begin{equation}
q =\rho_+^2 d\theta^2 +  \frac{(r_+^2 + a^2)^2 \sin^2\theta}{\rho_+^2} d\phi^2
\end{equation} where $\rho_+ = \rho(r_+, \theta)$, and has scalar curvature
\begin{equation}
R_q = \frac{2(r_+^2 + a^2)(4r_+^2 - 3 \rho_+^2)}{\rho_+^3}.
\end{equation}

A straightforward computation gives the stability operator:
\begin{equation}\label{Kerr stab}
\begin{aligned}
		L(a) \psi  =& -\frac{1}{\sin\theta} \frac{\p}{\p\theta} \left( \frac{\sin\theta}{\rho_+^2}\frac{\p\psi}{\p\theta} \right) - \frac{\rho^2_+}{(r_+^2 + a^2)^2 \sin^2\theta} \left(\frac{\p^2 \psi}{\p\phi^2} - 2 X_\phi \frac{\p\psi}{\p\phi} \right) \\ 
		&+ \left(\frac{1}{4r_+^2} - \frac{a^2}{(r_+^2 + a^2)^2}  + \frac{3 r_+^2 (r_+^2 + a^2)}{\rho^6_+}  - \frac{2a^2 + 3r_+^2}{\rho_+^4}  + \frac{(r_+^2 - a^2) (3r_+^2 + a^2)}{4r_+^2 (r_+^2 + a^2) \rho^2_+} \right) \psi ,
\end{aligned}
\end{equation}
where $X=X_\phi d\phi$ is given by 
\begin{equation}
X = \frac{ M a \sin^2 \theta ( 3r_+^4 + a^2 r_+^2 ( 1 + \cos^2\theta) - a^4 \cos^2\theta)}{\rho^4_+ (r_+^2 + a^2)} d\phi.
\end{equation} When $a=0$, \eqref{Kerr stab} reduces simply to \eqref{LSchw}.
Note that \eqref{Kerr stab} depends on three parameters $(M,a,r_+)$ satisfying $r_+^2 - 2M r_+ + a^2 = 0$. Here we have written everything just in terms of $a$ and $r_+(a)$, so the $a$-dependence of $M$ is hidden in the function $r_+$. This form covers all cases we are interested in, since any curve $s \mapsto (M(s), a(s), r_+(s))$ in the parameter space for which $\frac{da}{ds}(0) \neq 0$ can be parameterized by $a$, and we are only interested in perturbations that change the angular momentum. (For a variation in $M$ and $r_+$, with $a=0$ fixed, the spectrum is trivially obtained from \eqref{spec:unperturbed}.)

\section{Degenerate perturbation theory:  proof of Theorem \ref{thm:HF}}
We now compute the first variation of the degenerate eigenvalue $\lambda_\ell$. To motivate the proof, if we \emph{assume} the eigenvalues of $L(a)$ near $\lambda_\ell$ can be arranged into analytic branches $\lambda_\ell^m(a)$, an easy calculation (which is given below) shows that the derivatives $(\lambda_\ell^m)'(0)$ are precisely the eigenvalues of the Hellmann--Feynman matrix $T$ defined in \eqref{Tdef}. To make this rigorous, we use a result from perturbation theory which guarantees the existence of such analytic branches $\lambda_\ell^m(a)$ provided: 1) $L(a)$ is analytic; 2) $\lambda_\ell$ is a semi-simple eigenvalue of $L(0)$; and 3) the matrix $T$ has distinct eigenvalues. The first condition follows easily from the definition of $L(a)$, and the second is immediate because $L(0)$ is self-adjoint. The third condition will be verified below by an explicit computation of the matrix $T$.

We view $L(a)$ as an unbounded operator on the Hilbert space $L^2(S^2)$, with domain $H^2(S^2)$ independent of $a$. Using \eqref{Kerr stab}, we see that for each function $\psi \in H^2(S^2)$, the map $a \mapsto L(a)\psi \in L^2(S^2)$ is analytic in a neighbourhood of $a=0$, and hence $L(a)$ is an analytic family. (Using the terminology of Kato, it is a holomorphic family of type (A); see \cite{K76} for definitions and discussion.)

In general, this is not enough to conclude that the eigenvalues and eigenfunctions depend analytically (or even differentiably) on $a$. For instance, the eigenvalues and eigenvectors of the analytic matrix family $A(a) = \left( \begin{smallmatrix} 0 & 1 \\ a & 0 \end{smallmatrix} \right)$ are continuous but not differentiable at $a=0$. The problem here is that the eigenvalue $0$ of $A(0)$ is not semi-simple, i.e. its algebraic and geometric multiplicities do not coincide.

Since $L(0)$ is self-adjoint, all of its eigenvalues are semi-simple. It follows from \cite[Theorem II-2.3]{K76} that the eigenvalues of $L(a)$ in a small neighbourhood of $\lambda_\ell$, of which there are $2\ell + 1$ (counted with multiplicity), can be represented by continuously differentiable functions of $a$, whose derivatives are given by the eigenvalues of the operator $PL'(0)P\big|_V$, where $P$ denotes orthogonal projection onto the eigenspace $V := \ker(L(0) - \lambda_\ell)$.

The self-adjointness of $L(0)$ is still not enough to guarantee higher differentiability of the eigenvalues. However, it turns out the eigenvalues of $PL'(0)P\big|_V$ are distinct (they are precisely the values appearing in \eqref{eq:lambdalm}, as will be seen below), and this is enough to get analyticity of the eigenvalues and eigenfunctions; see \cite[\S 7.3.2]{B85} or \cite[\S II-2.3]{K76} for an in depth discussion.

The eigenspace $V := \ker(L(0) - \lambda_\ell)$ is spanned by spherical harmonics $Y_\ell^m$, $-\ell \leq m \leq \ell$. In this basis we can represent the operator $PL'(0)P \big|_V$ by a matrix, which we denote $T$. It has components
\begin{equation}\label{Tdef}
	T_{mn} = \left<P L'(0) P Y_\ell^n, Y_\ell^m \right>_{L^2(S^2)} 
	= \int_{S^2} \big(L'(0) Y_\ell^n \big)\overline{Y_\ell^m}.
\end{equation}

Before proving the theorem, we give a formal calculation to motivate the definition of the matrix $T$ in \eqref{Tdef}. Let $\lambda(a)$ denote an eigenvalue curve, with corresponding eigenfunction $\psi(a)$, and assume that both are differentiable at $a=0$. Differentiating the eigenvalue equation $L(a) \psi(a) = \lambda(a) \psi(a)$ and setting $a=0$, we have
\[
	L'(0) \psi(0) + L(0) \psi'(0) = \lambda'(0) \psi(0) + \lambda(0) \psi'(0).
\]
Multiplying by $\overline{Y_\ell^m}$ and integrating over $S^2$, we obtain
\begin{equation}\label{eq:HF1}
	\int_{S^2} \overline{Y_\ell^m} \big(L'(0) \psi(0)\big) = \lambda'(0) \int_{S^2} \overline{Y_\ell^m} \psi(0),
\end{equation}
where we have used the fact that $L(0)$ is self-adjoint.
Finally, since $\psi(0)$ is an eigenfunction for $\lambda_\ell$, it can be written as a linear combination of spherical harmonics, $\psi(0) = \sum_{-\ell}^\ell a_n Y_\ell^n$. Substituting this into \eqref{eq:HF1}, we find that
\[
	\sum_{n=-\ell}^\ell a_n \underbrace{\int_{S^2} \overline{Y_\ell^m} \big(L'(0) Y_\ell^n \big)}_{T_{mn}} = \lambda'(0) a_m.
\]
That is, the number $\lambda'(0)$ is an eigenvalue of the matrix $T$ defined in \eqref{Tdef}, with eigenvector $\bm{a} = (a_{-\ell}, \ldots, a_\ell)$.

We now give the proof of Theorem \ref{thm:HF}. Since $r_+(0) = 2\overbar M$ and $r_+'(0) = 0$, it follows easily from \eqref{Kerr stab} that
\[
	L'(0) = \frac{3}{8 \overbar M^3} \frac{\p}{\p\phi}.
\]
Since $\p_\phi Y_\ell^m(\theta,\phi) = im Y_\ell^m(\theta,\phi)$, we have
\[
	T_{mn} = \frac{3}{8 \overbar M^3} \int_{S^2} \overline{Y_\ell^m} (in Y_\ell^n)  = \frac{3ni}{8 \overbar M^3} \delta_{mn},
\]
so $T$ is a diagonal matrix with distinct eigenvalues. It now follows from \cite[Theorem 7.3.4 (p. 270)]{B85} that there exist analytic eigenvalue curves $\lambda_\ell^m(a)$, $-\ell \leq m \leq \ell$, with
\[
	\lambda_\ell^m(0) = \lambda_\ell, \qquad (\lambda_\ell^m)'(0) = \frac{3mi}{8 \overbar M^3} .
\]
Moreover, the eigenvector $\bm{a}^m = (a^m_{-\ell}, \ldots, a^m_\ell)$ of $T$ corresponding to the eigenvalue $\frac{3mi}{8 \overbar M^3}$ has entries $a^m_n = \delta_{mn}$, so for each $m$ we can find an analytic curve of eigenfunctions $\psi_\ell^m(a)$ for which
\[
	\psi_\ell^m(0) = \sum_{n=-\ell}^\ell a^m_n Y_\ell^n = Y_\ell^m.
\]
This completes the proof of Theorem \ref{thm:HF}.

\section{Symmetry reduction: proof of Theorem \ref{thm:2nd}}
We now calculate the second derivative of the eigenvalue $\lambda_\ell^0(a)$, which is constant to first order in $a$. Using a symmetry argument, we can reduce this to finding the derivative of a simple eigenvalue for a related Sturm--Liouville problem, with the reduced operator containing only even powers of $a$. This is a considerable simplification, since the general formula for the second derivative of an eigenvalue is rather involved even in the simple case (see \cite[eq. II-(2.36)]{K76}).

Dropping the $\phi$-dependent terms in $L(a)$, we obtain the Sturm--Liouville operator
\begin{equation}\label{Ltilde}
\begin{aligned}
		\widetilde L(a) \psi  =& -\frac{1}{\sin\theta} \frac{\p}{\p\theta} \left( \frac{\sin\theta}{\rho_+^2}\frac{\p\psi}{\p\theta} \right)  \\ 
		&+ \left(\frac{1}{4r_+^2} - \frac{a^2}{(r_+^2 + a^2)^2}  + \frac{3 r_+^2 (r_+^2 + a^2)}{\rho_+^6}  - \frac{2a^2 + 3r_+^2}{\rho_+^4}  + \frac{(r_+^2 - a^2) (3r_+^2 + a^2)}{4r_+^2 (r_+^2 + a^2) \rho_+^2} \right) \psi .
\end{aligned}
\end{equation}
This defines an unbounded, self-adjoint operator on the closed subspace $\widetilde L^2(S^2) \subset L^2(S^2)$ of functions not depending on $\phi$, with dense domain $H^2(S^2) \cap \widetilde L^2(S^2)$. As above, it is easily seen that $\widetilde L(a)$ is an analytic family of operators.


It follows immediately from the definition that every eigenfunction of $\widetilde L(a)$ is an eigenfunction of $L(a)$. Conversely, every $\phi$-independent eigenfunction of $L(a)$ is an eigenfunction of $\widetilde L(a)$. Therefore, $\lambda_\ell$ is a simple eigenvalue of $\widetilde L(0)$, corresponding to the eigenfunction $Y_\ell^0$, which does not depend on $\phi$. It follows that there is an analytic curve of eigenfunctions $\tilde\lambda_\ell(a)$ for $\widetilde L(a)$, with corresponding eigenfunctions $\tilde\psi_\ell(a) = Y_\ell^0 + \mathcal{O}(a)$. Comparing to Theorem \ref{thm:HF}, we conclude that $\tilde\lambda_\ell(a) = \lambda_\ell^0(a)$, and hence $\tilde\psi_\ell(a) = \psi_\ell^0(a)$. 

This means we can obtain the derivative of $\lambda_\ell^0(a)$ by differentiating $\tilde\lambda_\ell(a)$, which is simple. Moreover, from \eqref{Ltilde} we see that $\widetilde L(a)$ only depends on $a^2$, i.e. it contains no odd powers of $a$. This means $\widetilde L'(0) = 0$, and hence \cite[eq. II-(2.36)]{K76} implies
\begin{equation}\label{eq:lambda''}
	(\tilde\lambda_\ell)''(0) = \left<\widetilde L''(0) Y_\ell^0, Y_\ell^0 \right>_{L^2(S^2)},
\end{equation}
so it remains to compute the right-hand side. Our strategy is to write this in terms of integrals of products of associated Legendre polynomials, which can be calculated explicitly using Wigner 3-$j$ symbols. Some relevant properties of the 3-$j$ symbols are summarized in Appendix \ref{app:Wigner}.

We first recall that the spherical harmonics $Y_\ell^m$ are given by
\begin{equation}
	Y_\ell^m(\theta,\phi) = (-1)^m \sqrt{ \frac{2\ell+1}{4\pi} \frac{(\ell-m)!}{(\ell+m)!}} P_\ell^m(\cos\theta) e^{im\phi},
\end{equation}
with $P_\ell^m$ the associated Legendre polynomials. For $m=0$ these are just the Legendre polynomials $P_\ell$, and we have the relation
\begin{equation}
	\frac{\p}{\p\theta} P_\ell(\cos\theta) = P_\ell^1(\cos\theta).
\end{equation}

We now decompose $\widetilde L(a) = A(a) + B(a)$ as in \eqref{Ltilde}, where $A(a)$ is a second-order differential operator, and $B(a)$ is zeroth order. For the first term we integrate by parts to find
\begin{align*}
	\left< A(a) Y_\ell^0, Y_\ell^0 \right>_{L^2(S^2)} &= - \int_0^{2\pi} \int_0^\pi \frac{1}{\sin\theta} \frac{\p}{\p\theta} \left( \frac{\sin\theta}{\rho_+^2}\frac{\p Y_\ell^0}{\p\theta} \right) \overline{Y_\ell^0} \sin\theta\,d\theta d\phi \\
	&= 2\pi \int_0^\pi \frac{\sin\theta}{\rho_+^2}\frac{\p Y_\ell^0}{\p\theta} \frac{\p \overline{Y_\ell^0}}{\p\theta} \,d\theta \\
	&= \frac{2\ell + 1}{2} \int_0^\pi \frac{\sin\theta}{\rho_+^2}P_\ell^1(\cos\theta)^2 \,d\theta
\end{align*}
and hence
\begin{align*}
	\left< A''(0) Y_\ell^0, Y_\ell^0 \right>_{L^2(S^2)} 
	&= - \frac{2\ell+1}{16 \overbar M^4} \int_0^\pi \big(2 M r_+''(0) + \cos^2\theta \big) P_\ell^1(\cos\theta)^2 \sin\theta\, d\theta \\
	&= \frac{2\ell+1}{48 \overbar M^4} \int_{-1}^1 \big(P^2_2(z) - 6M r_+''(0) - 3 \big) P_\ell^1(z)^2 \, dz,
\end{align*}

where we have made the substitution $z = \cos\theta$ and used the fact that $P_2^2(z)= 3(1-z^2)$. Using the orthogonality relation
\[
	\int_{-1}^1 P_\ell^1(z)^2 \, dz = \frac{2\ell(\ell+1)}{2\ell+1}
\]
and the identity
\[
	\int_{-1}^1 P^2_2(z) P_\ell^1(z)^2 \, dz = \frac{12 \ell^2(\ell+1)^2}{(2\ell+3)(2\ell+1)(2\ell-1)}
\]
from the appendix, we find that
\begin{equation}\label{eq:Afinal}
	\left< A''(0) Y_\ell^0, Y_\ell^0 \right>_{L^2(S^2)} 
	= \frac{1}{4\overbar M^4} \frac{\ell^2(\ell+1)^2}{(2\ell+3)(2\ell-1)} - \frac{2\overbar M r_+''(0) + 1}{8 \overbar M^4} \ell(\ell+1).
\end{equation}

Similarly, for the zeroth-order part we compute
\[
	B''(0) = -\frac{1}{32 \overbar M^4} \big(8\overbar M r_+''(0) + 5 + 15 \cos^2\theta \big)
\]
and so
\begin{align*}
	\left<B''(0) Y_\ell^0, Y_\ell^0 \right>_{L^2(S^2)} &= -\frac{\pi}{16 \overbar M^4} \int_0^\pi \big(8\overbar M r_+''(0) + 5 + 15 \cos^2\theta \big) \big|Y_\ell^0 \big|^2 \sin\theta\, d\theta \\
	&= - \frac{2\ell + 1}{32 \overbar M^4} \int_{-1}^1 \big(4\overbar M r_+''(0) + 5 + 5 P_2(z) \big) P_\ell(z)^2\, dz.
\end{align*}
Using the orthogonality relation
\[
		\int_{-1}^1 P_\ell(z)^2 \, dz = \frac{2}{2\ell + 1}
\]
and the identity
\[
	\int_{-1}^1 P_2(z) P_\ell(z)^2 \, dz = \frac{2\ell(\ell + 1)}{(2\ell+3)(2\ell+1)(2\ell-1)}
\]
from the appendix, we find that
\begin{equation}\label{eq:Bfinal}
	\left<B''(0) Y_\ell^0, Y_\ell^0 \right>_{L^2(S^2)} = - \frac{4\overbar M r_+''(0) + 5}{16 \overbar M^4} - \frac{5}{16 \overbar M^4} \frac{\ell(\ell+1)}{(2\ell+3)(2\ell-1)}.
\end{equation}

Adding \eqref{eq:Afinal} and \eqref{eq:Bfinal} and substituting into \eqref{eq:lambda''} completes the proof.

\appendix

\section{Wigner 3-$j$ symbols}
\label{app:Wigner}
Here we review some properties of the Wigner 3-$j$ symbols (as used above in the proof of Theorem \ref{thm:2nd}), following the presentation of \cite[Appendix C]{M65}. It is well known that the integral of three Legendre polynomials can be written in terms of Wigner 3-$j$ symbols as
\begin{equation}\label{eq:3LP}
	\int_{-1}^1 P_{\ell_1}(z) P_{\ell_2}(z) P_{\ell_3}(z)\,dz = 2
	\begin{pmatrix} \ell_1 & \ell_2 & \ell_3 \\ 0 & 0 & 0 \end{pmatrix}^2.
\end{equation}
In general the 3-$j$ symbols are difficult to compute explicitly, but the following special case
\begin{equation}\label{eq:ll2}
	\begin{pmatrix} 2 & \ell & \ell \\ 0 & 0 & 0 \end{pmatrix} = (-1)^{\ell+1} \sqrt{\frac{\ell(\ell+1)}{(2\ell+3)(2\ell+1)(2\ell-1)}}
\end{equation}
is easily obtained from \cite[Eq. (C.23b)]{M65}, so we have
\begin{equation}
	\int_{-1}^1 P_2(z) P_\ell(z)^2\,dz = \frac{2\ell(\ell + 1)}{(2\ell+3)(2\ell+1)(2\ell-1)}.
\end{equation}

More general (and complicated) formulas exist for integrals of associated Legendre polynomials. For $m_3 = m_1 + m_2$ we have
\begin{align}\label{eq:3ALP}
\begin{split}
	\int_{-1}^1 P_{\ell_1}^{m_1}(z) P_{\ell_2}^{m_2}(z) P_{\ell_3}^{m_3}(z)\,dz = 2 (-&1)^{m_3}
	\sqrt{ \frac{(\ell_1 + m_1)!(\ell_2 + m_2)!(\ell_3 + m_3)!}{(\ell_1 - m_1)!(\ell_2 - m_2)!(\ell_3 - m_3)!}} \\
	&\times \begin{pmatrix} \ell_1 & \ell_2 & \ell_3 \\ 0 & 0 & 0 \end{pmatrix}
	\begin{pmatrix} \ell_1 & \ell_2 & \ell_3 \\ m_1 & m_2 & -m_3 \end{pmatrix},
\end{split}
\end{align}
see \cite[eq. (30)]{DL02}. Choosing $m_1 = m_2 = m_3 = 0$, we get \eqref{eq:3LP} as a special case. The other case we need is $\ell_1 = \ell_2 = \ell$, $\ell_3 = 2$, $m_1 = m_2 = 1$ and $m_3 = 2$. Using
\begin{equation}
	\begin{pmatrix} \ell & \ell & 2 \\ 1 & 1 & -2 \end{pmatrix} = (-1)^{\ell+1} \sqrt{\frac32} \sqrt{\frac{\ell(\ell+1)}{(2\ell+3)(2\ell+1)(2\ell-1)}}
\end{equation}
together with \eqref{eq:ll2}, we find that
\begin{equation}
	\int_{-1}^1 P_2^2(z) P_{\ell}^1(z)^2 \,dz = \frac{12 \ell^2(\ell+1)^2}{(2\ell+3)(2\ell+1)(2\ell-1)}.
\end{equation}

\bibliographystyle{amsplain}
\bibliography{MOTS}

\end{document}